\documentclass[]{spie}  

\usepackage{amsmath,amsfonts,amssymb}
\usepackage{graphicx}
\usepackage[colorlinks=true, allcolors=blue]{hyperref}

\title{Integrated coronagraphy and wavefront sensing with the PIAACMC}

\author[a]{Sebastiaan Y. Haffert}
\author[a]{Jared R. Males}
\author[a,b,c,d]{Olivier Guyon}
\affil[a]{University of Arizona, Steward Observatory, Tucson, Arizona, United States}
\affil[b]{Wyant College of Optical Science, University of Arizona, 1630 E University Blvd, Tucson, AZ 85719, USA}
\affil[c]{Astrobiology Center, National Institutes of Natural Sciences, 2-21-1 Osawa, Mitaka, Tokyo, JAPAN}
\affil[d]{National Astronomical Observatory of Japan, Subaru Telescope, National Institutes of Natural Sciences, Hilo, HI 96720, USA}

\authorinfo{Further author information: (Send correspondence to S.Y.H)\\A.A.A.: E-mail: shaffert@arizona.edu}

\begin{document} 
\maketitle

\begin{abstract}
Uncorrected wavefront errors create speckle noise in high-contrast observations at small inner-working angles. These speckles can be sensed and controlled by using coronagraph integrated wavefront sensors. Here, we will present how the Phase Induced Amplitude Apodized Complex Mask Corongraph (PIAACMC) can be integrated with both a Self-Coherent Camera (SCC) for focal plane wavefront sensing and an extremely sensitivity high-order pupil plane Zernike wavefront sensor (ZWFS). Non-common path aberrations can be completely erased by integrating both sensors into the PIAACMC, which is of extremely high importance in high-contrast imaging. 
\end{abstract}


\section{INTRODUCTION}
\label{sec:intro}
Upcoming ground-based telescopes like the Giant Magellan Telescope (GMT) and the European Extremely Large Telescope (E-ELT) have direct imaging of Earth-like planets around M stars as one of their major science goals \cite{kasper2021pcs, males2022conceptual}. The next flagship mission from NASA, the Habitable Worlds Observatory (HWO), likewise has direct imaging of Earth-like planets, although around Sun-like stars, as the number one priority \cite{national2021pathways}. The most important optical component within high-contrast instruments is the coronagraph. The coronagraph removes light from any on-axis object will passing through off-axis light, such as light from exoplanets. The goal of the coronagraph in space-like conditions is to reach a contrast of $10^{-10}$ over a bandwidth of 20\% at a separation of a couple times the diffraction limit. Ground-based coronagraphs will not need such strict contrast requirements because residual atmospheric turbulence after adaptive optics creates a contrast floor on the level of $\sim10^{-6}$ to $\sim10^{-4}$ \cite{guyon2018exao}. Most coronagraphs have been designed for monolithic mirrors. However, all of the future telescopes will have complex segmented apertures. Coronagraph designs for segmented apertures show a gap in performance to those for monolithic mirrors, this is, however, not a limit set by physics but by a technology gap \cite{belikov2021theoretical}.

One promising coronagraph architecture is that of the Phase Induced Amplitude Apodization family \cite{guyon2003phase, guyon2010high, guyon2013high}. The PIAA coronagraphs follow a classic Lyot-style architecture with an entrance apodization, a focal plane mask and a Lyot stop. Classic Lyot coronagraphs usually use grayscale transmission masks, which results in lower planet throughput and increased inner-working angle. The unique aspect of the PIAA corongraphs is that they use a set of aspheric optics to loselessy apodize the incoming pupil by redistributing the light across the aperture. This creates an effective apodization of the pupil which suppresses the rings of the Airy diffraction pattern of the stellar Point Spread Function (PSF) \cite{guyon2003phase}. An extension of the PIAA coronagraph uses complex focal plane masks that manipulate the light in both amplitude and phase. This resulted in the PIAA Complex Mask Coronagraph (PIAACMC) \cite{guyon2013high}.

\begin{figure}
    \begin{center}
        \includegraphics[width=\textwidth]{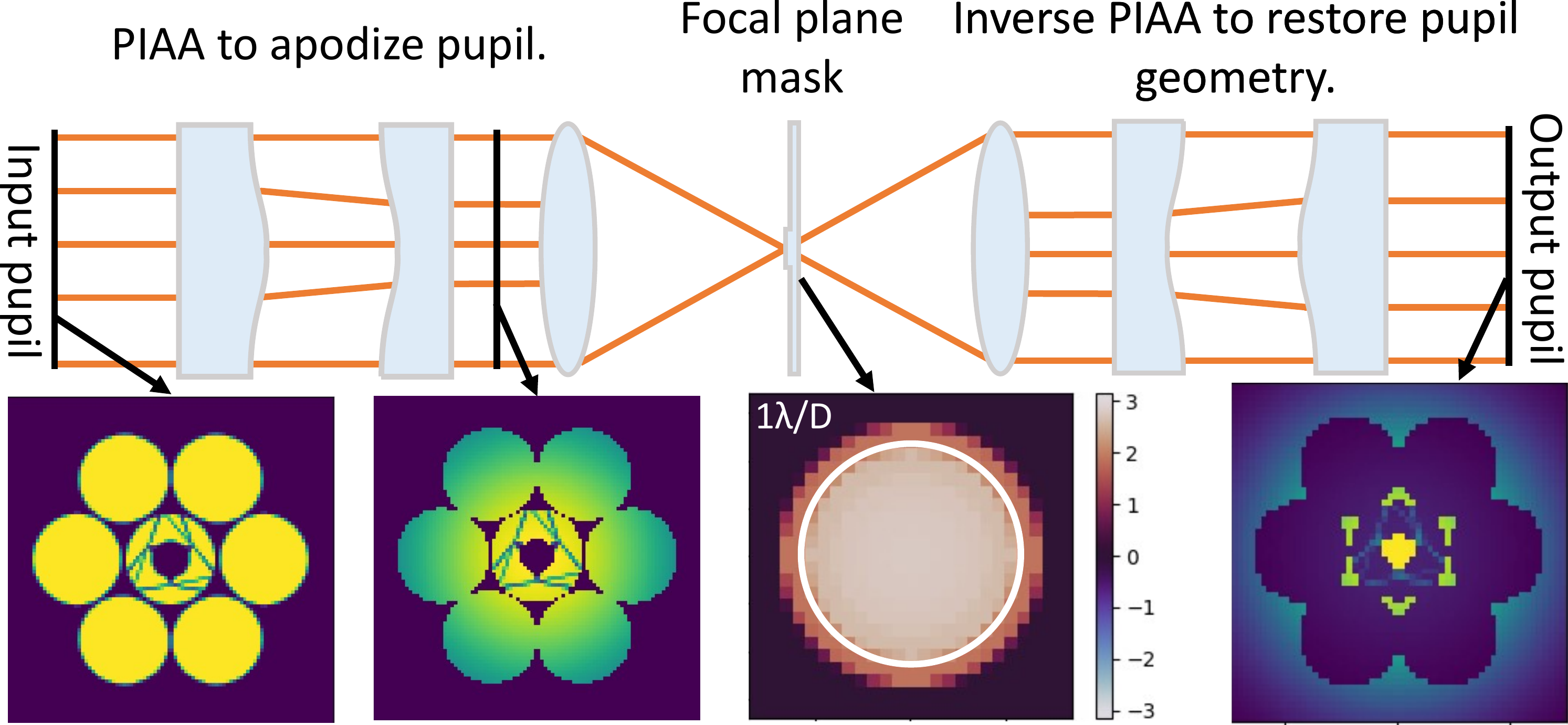}
	\end{center}
    \caption{The Phase Induced Amplitude Apodized Complex Mask Coronagraph architecture. The incoming beam is apodized losslessly by a set of aspheric optics to match the intrinsic response of the focal plane mask. The partially phase-shifting mask will then destructively interfere all light inside the geometric pupil. It is necessary to use a second set of PIAA lenses in reverse to restore the original pupil geometry and regain field of view.}
    \label{fig:piaacmc} 
\end{figure}

The benefit of the PIAACMC is that it provides a very small inner-working angle (IWA$\sim 1\lambda/D$) and high planet throughput ($>70\%$) \cite{guyon2010high}. The PIAACMC architecture works well for both ground- and space-based telescopes; it can be designed to reach the required contrast levels for imaging Earth-like planets ($\sim10^{-10}$) because it is robust against complicated telescope pupil architectures which include segment gaps, spiders, and a secondary obscuration. The primary reason for robustness comes from the aperture apodization. Typical focal plane mask sizes are 1 to 2 $\lambda/D$ in diameter. The original design was closer to 1 $\lambda/D$ \cite{guyon2013high}. The incoming beam is spatially filtered by the focal plane mask which creates a reference beam. The PIAACMC can then be described as the sum of two electric fields; the original incoming wavefront and the reference beam. The PSF has an almost uniform shape within a diameter of 1 $\lambda/D$ regardless of the shape of the aperture. This cause all telescopes to have nearly identical reference beams. The PIAA lenses are then used to apodize the aperture in such a way that the amplitude profile of the pupil matches the profile of the reference beam. A pi-phase shifting focal plane mask can then create perfect destructive interference. This is the key point of the PIAACMC. This specific design only works monochromatically because the PSF scales with wavelength due to diffraction. Multi-zone phase-shifting masks have been designed to reach deep contrast over broad spectral bandwidths \cite{knight2018phase, martinez2020design}. An example of the PIAACMC architecture is shown in Figure \ref{fig:piaacmc}.

The PIAACMC can achieved a small IWA, high throughput and a deep null which comes at a cost of high sensitivity against wavefront errors. These wavefront errors will create a strong leakage through the coronagraph if they are not controlled. Integrating wavefront sensors into the coronagraph can alleviate these concerns. There are multiple advantages to this approach:
\begin{itemize}
    \item The wavefront sensor measures the wavefront errors where they matter. It has been shown that aberrations upstream of the coronagraph matter more than aberrations that are downstream of the coronagraph.
    \item An integrated approach minimizes the amount of optics and therefore the chances for additional non-common path aberrations.
    \item The aberrations will be measured simultaneously with the science observations, so they are non-destructive.
    \item There is potential to use the wavefront sensor telemetry for Coherence Differential Imaging (CDI).
\end{itemize}

Therefore, it is crucial to investigate how to integrate wavefront sensors into the coronagraph. We discuss how two different wavefront sensors can be integrated into the PIAACMC coronagraph. We discuss how the PIAACMC is a natural Zernike wavefront sensor, which enables the retrieval of high-order aberrations from pupil plane images. The second wavefront sensor is the Self Coherent Camera, which creates a common path Fizeau interferometer out of the coronagraphic image.

\section{Zernike PIAACMC}
Interferometers are very sensitive wavefront sensors and they are not blind to certain modes that are plaguing current system such as the low-wind effect or differential piston. However, the interferometer should be a common path interferometer to remove the effects of any differential aberrations between the reference beam and the sampled beam. A second requirement is that all light should be used, no light can be blocked. This maximizes the throughput and will enable use to use all available photons. The ZWFS \cite{n2013calibration} is one of the few sensors that meets these requirements. The ZWFS uses a phase shifting mask that shifts the phase of the core of the Point Spread Function (PSF). The phase-shifted core creates the reference beam for the measurement. It is near optimal in terms of robustness against photon noise and read noise \cite{chambouleyron2021variation,chambouleyron2022optimizing}. The sensitivity and accuracy of the ZWFS has been demonstrated by picometer precision in wavefront errors \cite{steeves2020picometer}.

\begin{figure}
    \begin{center}
        \includegraphics[width=\textwidth]{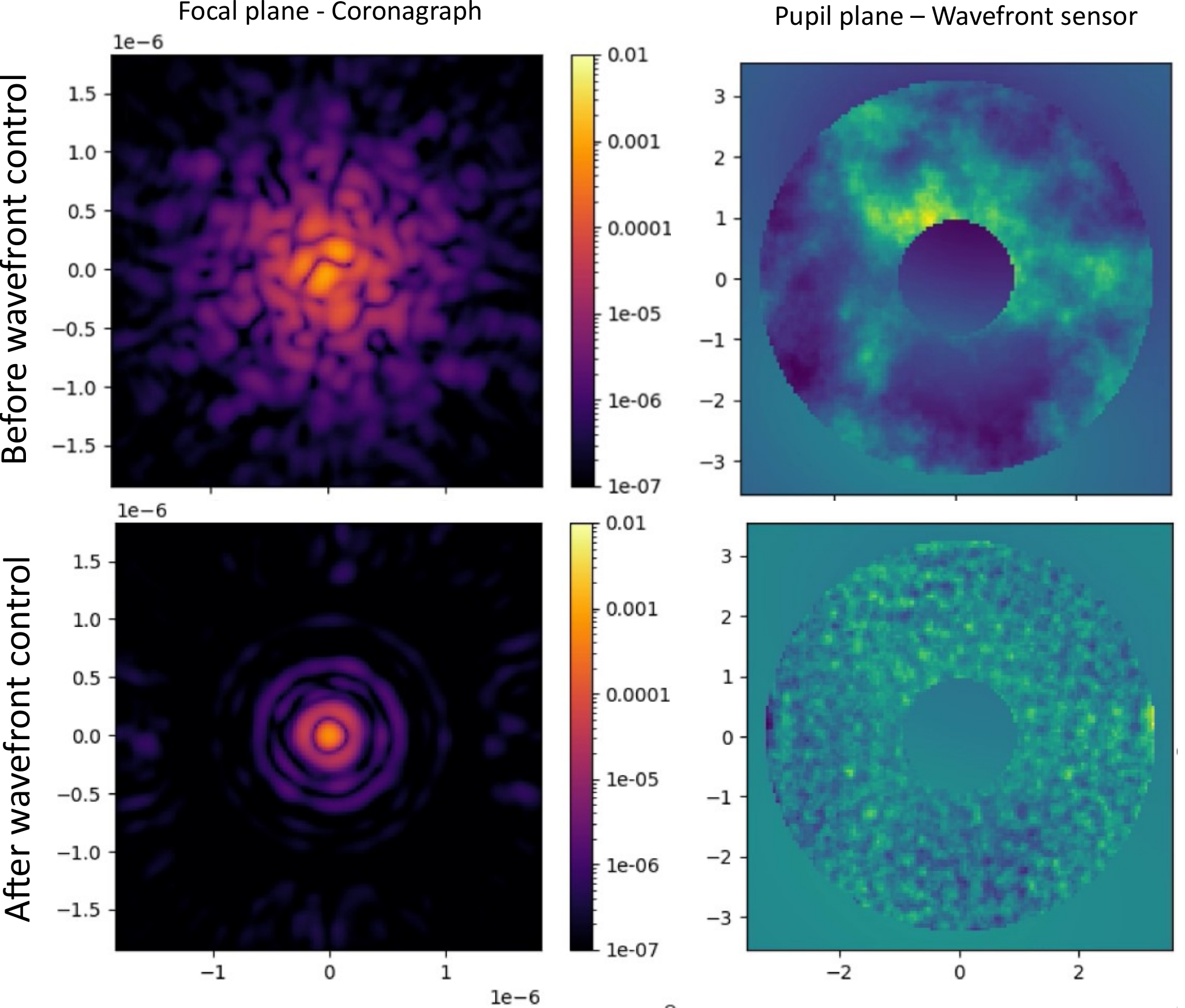}
	\end{center}
    \caption{Metamaterials are engineered materials that use sub-wavelength structures. The exact properties depend on the specific substructure. So, the electric field can be manipulated on a point-by-point basis by changing the substructure. We propose an integrated wavefront sensor and coronagraph by using metamaterials that have a different response for the two polarizations. In this way we can combine the most sensitive wavefront sensor (PIAA-ZWFS) with an extremely high performance and small inner-working angle coronagraph (PIAA-CMC). The figures on the right show such a design in action.}
    \label{fig:piaazwfs} 
\end{figure}

The ZWFS has a nearly identical layout to that of the PIAACMC coronagraph. The only difference is the size of the phase step of the focal plane mask. The PIAACMC uses a $\pi$ phase step to destructively interfere all light and the ZWFS uses a $\pi/2$ phase step to optimally probe the wavefront \cite{chambouleyron2021variation}. Therefore, it is almost natural to integrate the PIAACMC with a Zernike wavefront sensor. The only question is how to impart the difference in phase. We propose two solutions. The first uses out-of-band light for the wavefront sensing. The light within the spectral band of the coronagraph sees an optimized focal plane mask for destructive interference. However, light that comes from outside of the coronagraphic bandpass will not get the same phase shift. This means that the out-of-band light will behave more similar to a ZWFS. Such a system has been simulated in Figure \ref{fig:piaazwfs}. The in-band light is used for the coronagraphic observations and the out-of-band light is used for wavefront sensing to stabilize the dark hole. This is an useful architecture because it does not throw away any light from the science channel.

\section{PIAACMC with a Self Coherent Camera: FAST-PIAACMC}
The Self Coherent Camera (SCC) is an integrated coronagraph and WFS that uses a high-frequency spatial modulation of the stellar speckles to estimate the complex electric field \cite{baudoz2006scc}. An off-axis pinhole in the Lyot stop of the coronagraph is used to create the spatial modulation. An on-axis source will hit the coronagraph focal plane mask and will then diffract light outside outside the geometric pupil and therefore into the reference hole. This creates Fizeau fringes only for the on-axis source, which can be used to measure the incoming wavefront in both phase and amplitude. However, the reference hole has to be placed at least 1 pupil diameter from the pupil edge to reconstruct the electric field from a single measurement \cite{galicher2010scc}.

The SCC was originally developed and optimized for space-based observatories \cite{baudoz2006scc,galicher2010scc}. Recent work has proven that it can also be used on ground-based telescopes \cite{galicher2019sccpalomar,singh2019sccaohalo}. The main challenge for the SCC for ground-based telescopes is the weak intensity modulation. In space-based conditions the coronagraphic dark hole can reach contrast levels of $10^{-8}$ to $10^{-10}$ which matches with the created modulation amplitude. Ground-based telescopes on the other hand have a post-coronagraphic contrast of $10^{-3}$ to $10^{-6}$, which is several orders of magnitude higher in amplitude.

The Fast Atmospheric SCC Technique (FAST) modifies the coronagraphic focal plane mask to increase the throughput at the off-axis pinhole \cite{gerard2018fast}. FAST can run with a much shorter exposure time or equivalently, work on much fainter objects due to the increased amount of light in the reference pinhole. Another way to increase the throughput through the pinhole is to place it closer to the edge of the pupil. This increases the throughput but requires an unmodulated image to reveal the modulation fringes. The modulated SCC can be implemented in several ways \cite{martinez2019fmscc, bos2021polarization, haffert2022spectrally}. Here we create a FAST-PIAACMC, which can be naturally built into the focal plane mask.

\begin{figure}
    \begin{center}
        \includegraphics[width=\textwidth]{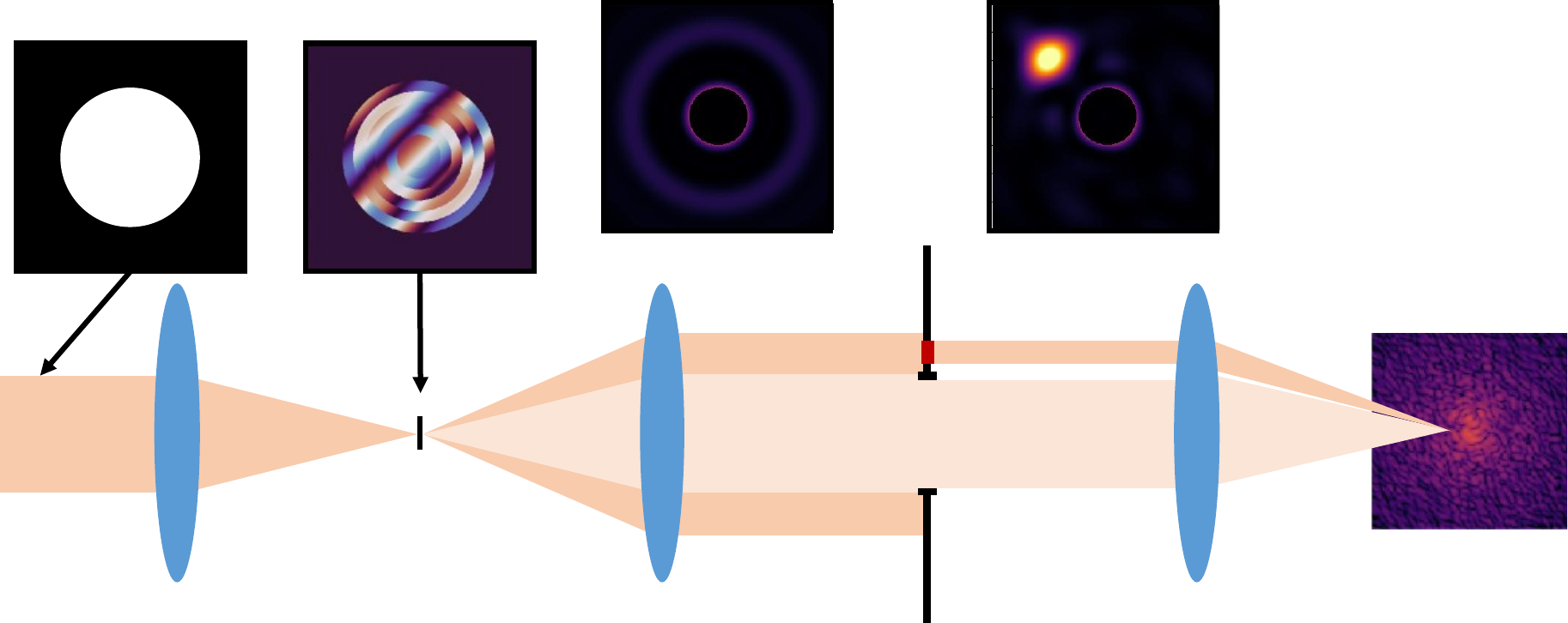}
	\end{center}
    \caption{The beam is focused onto a focal-plane mask that will diffract on-axis light outside of the geometric pupil. The diffracted light illuminates an off-axis pinhole which creates Fizeau fringes in the focal plane. The normal PIAACMC has very little light at the position of the pinhole with low SNR on the fringes consequently. The FAST-PIAACMC optimizes the focal plane mask to suppress the light inside the Lyot stop and to increase the throughput at the position of the pinhole, which increases SNR on the fringes. This allows the FAST-PIAACMC to be used with short enough exposure times to freeze atmospheric speckles.}
    \label{fig:fastpiaacmc} 
\end{figure}

The FAST-PIAACMC is made by optimizing the focal plane mask not only for contrast, but also for throughput through the off-axis pinhole. Optimizing the focal plane mask with just zones did not lead to solutions that increased the amount of light in the pinhole \cite{knight2018phase, martinez2020design}. The off-axis light was created by creating circular zones that are modulated by cosine phase patterns. The phase pattern is then,
\begin{equation}
\phi = \sum_i m_i a_i \cos{\left(2\pi f_i x + \psi_i\right)}.
\end{equation}
Here, $\phi$ is the focal plane phase pattern, $m_i$ the binary mask defining a particular zone in the focal plane, $a_i$ the amplitude of the cosine modulation, $f_i$ the frequency of the cosine and $\psi_i$ the phase offset of the cosine. There are several advantages to optimizing this pattern. The first is that a cosine grating will create off-axis copies of the spatially filtered pupil exactly at the place of the pinhole. A second advantage is that the cosine gratings will scatter away light from the focal plane mask effectively creating an amplitude apodization. This makes the PIAACMC more efficient because it needs some amount of amplitude apodization, which is the complex mask part of the coronagraph. Using gratings replaces the amplitude mask with a phase only solution. This solves two issues at the same time; it increases the pinhole throughput and gives us control over amplitude without needing a second optic. Figure \ref{fig:fastpiaacmc} shows the optimized FAST-PIAACMC mask and also shows the difference in intensity at the position of the pinhole.

\begin{figure}
    \begin{center}
        \includegraphics[width=\textwidth]{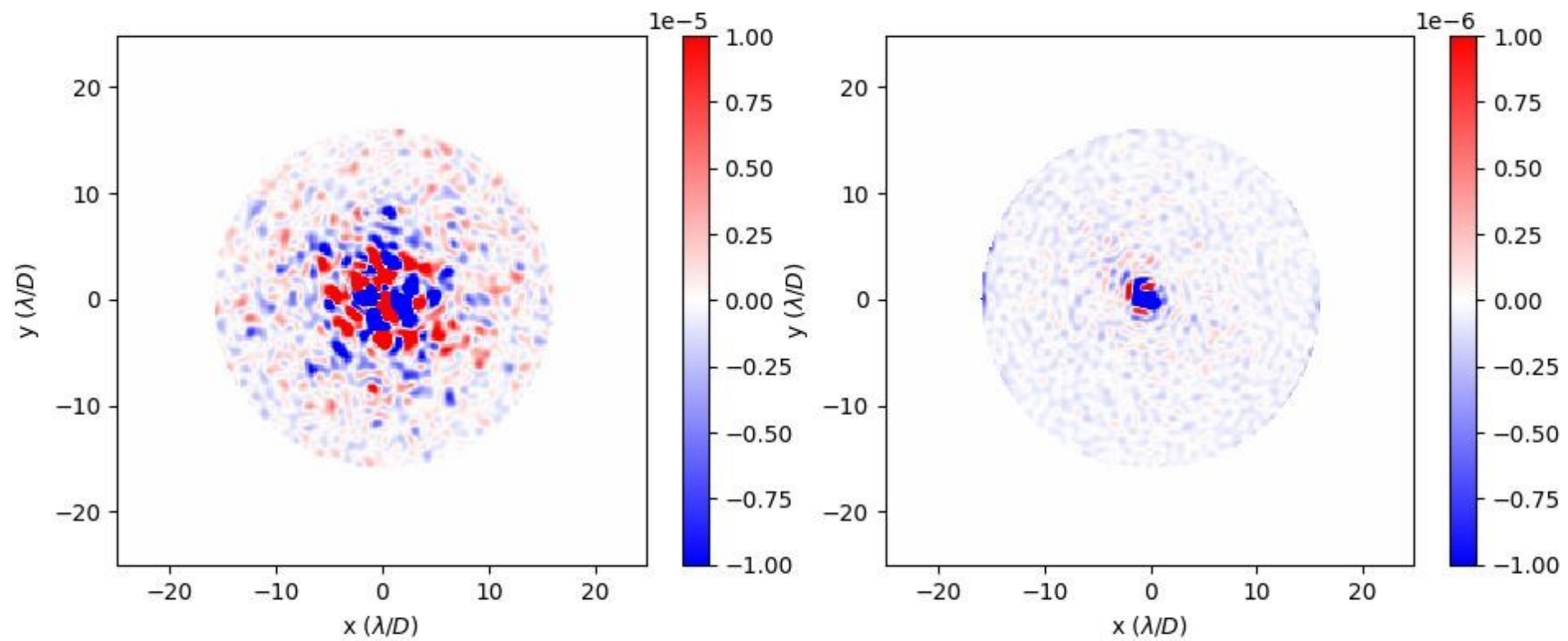}
	\end{center}
    \caption{The SCC measures the full electric field for each single frame. This can be used for both focal-plane wavefront control and Coherence Differential Imaging (CDI). The residuals are representative of what GMT could reach within 10 seconds.}
    \label{fig:fastpiaacmc_result} 
\end{figure}

Exoplanet light is incoherent with star light. Therefore, we can discriminate between planet and starlight if we can separate light into coherent and incoherent light. This post-processing method is called Coherence Differential Imaging (CDI). CDI is a natural post-processing technique for the SCC because the SCC instantaneously measures the electric field of the stellar speckles allowing us to easily disentangle coherent and incoherent light \cite{galicher2010scc}. Figure \ref{fig:fastpiaacmc_result} shows the results of running the FAST-PIAACMC in closed loop to remove stellar speckles and using CDI as a post-processing technique. The simulations have been done for a GMagAO-X like system for GMT. The results show that we could reach a post-processed sensitivity of $\sim 10^{-7}$ with only several seconds of exposure time.

\section{Conclusion}
We have shown how the PIAACMC can be integrated with different wavefront sensors for NCPA control. The ZWFS and FAST-PIAACMC use different approaches to achieve wavefront sensing. An interesting question is if both the ZWFS and FAST-PIAACMC can be integrated into a single because. Making both the pupil and focal plane able to sense the electric field breaks the typical information loss between the planes. This means we use the PIAACMC-ZWFS telemetry to estimate the speckle pattern. The ZWFS has nearly perfect sensitivity and does not throw away light which means that it will measure the wavefront and therefore the speckles with the highest possible signal to noise. Any differential error between the focal plane and pupil can also be included into the model because the FAST-PIAACMC will be able to measure those. This combination will create an opportunity to reach the fundamental photon noise sensitivity limit which allows us to image Earth-like planets with the next generation of telescopes.

\acknowledgments 
Support for this work was provided by NASA through the NASA Hubble Fellowship grant \#HST-HF2-51436.001-A awarded by the Space Telescope Science Institute, which is operated by the Association of Universities for Research in Astronomy, Incorporated, under NASA contract NAS5-26555 and by the generous support of the Heising-Simons Foundation.

\bibliography{report} 
\bibliographystyle{spiebib} 

\end{document}